\documentclass[letter]{aa} 

\usepackage{graphicx}
\usepackage{txfonts}

\usepackage{silence}
\usepackage[flushleft]{threeparttable}
\usepackage{xcolor}
\usepackage{xspace}
\usepackage{longtable}
\usepackage{booktabs}
\usepackage{hyperref}
\usepackage{tabularx}

\newcommand{\jbu}{SN~2016jbu\xspace}
\newcommand{\ip}{SN~2009ip\xspace}
\newcommand{\bh}{SN~2015bh\xspace}
\newcommand{\bdu}{SN~2016bdu\xspace}
\newcommand{\gl}{SN~2005gl\xspace}
\newcommand{\lsq}{LSQ13zm\xspace}
\newcommand{\etacar}{$\eta$ Car\xspace}

\newcommand{\dusty}{\textsc{DUSTY}\xspace}
\newcommand{\bpass}{\textsc{BPASS}\xspace}

\newcommand{\msun}{$M_{\odot}$\xspace}
\newcommand{\kms}{$km~s^{-1}$\xspace}

\graphicspath{{./figures/}}

\begin{document} 
   \title{The impostor revealed: \jbu was a terminal explosion }
   \author{
   S.J. Brennan\inst{1}\thanks{Email: \url{sean.brennan2@ucdconnect.ie}}
   \and
   N. Elias-Rosa\inst{2,3}
   \and
   M. Fraser\inst{1}
   \and 
    S.D. Van Dyk\inst{4}
   \and
    J.D. Lyman \inst{5}
   }

    \institute{
    School of Physics, O’Brien Centre for Science North, University College Dublin, Belfield, Dublin 4, D04 V1W8, Ireland
    \and
    INAF -- Osservatorio Astronomico di Padova, vicolo dell'Osservatorio 5, Padova I-35122, Italy
    \and
    Institute of Space Sciences (ICE, CSIC), Campus UAB, Carrer de Can Magrans s/n, 08193 Barcelona, Spain
    \and 
    Caltech Spitzer Science Center, Caltech/IPAC, Mailcode 100-22, Pasadena, CA 91125, USA
    \and 
    Department of Physics, University of Warwick, Coventry CV4 7AL, UK
    }

   \date{\today}

 
  \abstract
    {In this Letter, we present recent observations from the Hubble Space Telescope of the interacting transient, \jbu , at +5 years. We find no evidence for any additional outburst from \jbu, and the optical source has now faded significantly below the progenitor magnitudes from early 2016. Similar to recent observations of \ip and \bh,  \jbu has not undergone a significant change in colour over the past 2 years, suggesting that there is a lack of on-going dust formation. We find \jbu is fading slower than that expected from radioactive nickel but faster than the decay of \ip. The late time lightcurve displays a non-linear decline and follows on from a re-brightening event that occurred $\sim$8 months after peak brightness, suggesting CSM interaction continues to dominate \jbu. While our optical observations are plausibly consistent with a surviving, hot, dust-enshrouded star, this would require an implausibly large dust mass. These new observations suggest that \jbu is  a genuine, albeit strange, supernova, and we discuss the plausibility of a surviving binary companion.}
    

   \maketitle
%

%
\section{Introduction}

The final moments of  massive stars' \citep[>~$8$~\msun; ][]{Woosley2002} lives are often not well understood. While the general picture of massive stars exploding as core-collapse SNe is well understood, it is increasingly apparent that some of these stars will experience enhanced mass loss before exploding \citep{Fraser2013,Yaron2017}.

In recent years, a group of transients has been classified, where it is uncertain whether these are core-collapse supernovae (CCSNe) or giant non-terminal eruptions. These events have been dubbed \ip-like transients and include such events as  \ip \citep{foley2011,Smith2010,Fraser2013,Pastorello2013,Mauerhan2013,Smith2014}, \bh \citep{Elias-Rosa2016,Thone2017,Boian2018,Jencson2022}, \bdu , \gl \citep{Pastorello2018}, \lsq \citep{Tartaglia2016}, and \jbu \citep{Kilpatrick2018,Brennan2021a,Brennan2021b}. 

These transients display several peculiar features that are challenging to explain with current stellar evolutionary theory. These include a history of erratic variability, followed by two luminous events (known as Event A and Event B), the latter reaching a magnitude comparable to Type IIn supernovae (SNe) \footnote{SNe showing signs of interaction with circumstellar material, with narrow emission lines seen in their spectra \citep{Schlegel1990,Filippenko1997}.} \citep[$\sim-18.5$~mag, ][]{Nyholm2020}, and a bumpy decline in their light curve. The late-time photometric evolution for these objects is slower than that expected for radioactive $^{56}$Ni decay \citep{Fox2015}. Additionally, these objects display smoothly evolving asymmetric Balmer emission lines, suggesting a complex, possibly disk-like, circumstellar material (CSM) \citep{Brennan2021b,Reilly2017}. Curiously, their ejected $^{56}$Ni mass is constrained to be relatively low (less than a few 0.01~\msun) \cite{Smith2014,Margutti2014,Brennan2021b}. This has led some authors to consider whether we are observing a core-collapse event unfold, or a giant stellar eruption, perhaps similar to the non-terminal Giant Eruption of \etacar, i.e. a SN Impostor \citep{VanDyk2012,Pastorello2013,Elias-Rosa2016,Hirai2021}.

\ip has recently been re-observed by \cite{Smith2022} using the Hubble Space Telescope (HST), who reported a source with a luminosity significantly lower than the progenitor observed in 1999 in $F606W$. Critically, \cite{Smith2022} find a constant colour for the late-time light curve of \ip, indicating an absence of dust formation. This strongly disfavours the possibility that the massive progenitor \citep[$\sim60 - 80$~\msun, ][]{Smith2010,foley2011} is being obscured by large amounts of newly-formed dust. Based on these recent observations, \cite{Smith2022} conclude that \ip was indeed a SN and will continue to fade. \cite{Jencson2022} have concluded a similar fate to \bh, where the light curve is now significantly fainter than the progenitor fading with a constant colour, ruling out a dust-enshrouded surviving star.

This Letter presents new late-time observations of \jbu. \jbu ( a.k.a Gaia16cfr ) offers a unique opportunity to search for a surviving star, as the progenitor was detected at multiple wavelengths, and was well constrained by \cite{Kilpatrick2018,Brennan2021b} to be a $\sim22$~\msun Yellow Hyper Giant (YHG).

Following \cite{Brennan2021a}, we take the distance modulus for NGC~2442 to be $31.60\pm0.06$~mag. We adopt a redshift of z=0.00489 and a Milky Way (MW) foreground extinction to be $A_V=0.556$~mag. We correct for foreground extinction using $R_V=3.1$ and the extinction law given by \citealp{Cardelli1989}.

\section{Observations}

We observed the site of \jbu in December 2021 during Cycle 29 with HST (ID: 16671, PI: N. Elias-Rosa) using the UV-Visible (UVIS) and Infrared (IR) channels of the Wide Field Camera 3 (WFC3/UVIS and WFC3/IR respectively). The objective of our proposal was to re-observe \jbu with the same filters that were used in early 2016 and 2019. These include WFC3/UVIS $F555W$, $F350LP$, $F814W$ and WFC3/IR $F160W$. Serendipitously, the host of \jbu, NGC 2442, was observed with WFC3/UVIS in $F275W$ in March 2021 (ID: 16287, PI: J. D. Lyman). Pipeline reduced images were downloaded from Mikulski Archive for Space Telescopes (MAST\footnote{\url{mastweb.stsci.edu/}}), and photometry was performed on these images using the {\sc dolphot} package \citep{2016ascl.soft08013D}. 

In all cases, images were masked for cosmic rays and other artefacts using the associated data quality files. Source detection was performed on a pipeline drizzled reference image, before PSF-fitting photometry was performed on the individual {\sc\_flt} /{\sc\_flc} frames. For the WFC3/UVIS data taken in Dec 2021, the pipeline drizzled $F350LP$ image was used as a reference frame for source detection; for the WFC3/IR data, the drizzled $F160W$ image was used as a reference. The 2021 March $F275W$ data were analysed separately, using the drizzled $F275W$ image as a reference.

Photometry for the point source at the position of \jbu is reported in Table. \ref{tab:hst_obs}. 

\begin{table}[!ht]
    \centering
    \caption{Observational log for HST + WFC3/UVIS and WFC3/IR images covering the site of \jbu from December 2021. We also include $F275W$ taken $\sim$9 months prior. Measured photometry (in the Vega-mag system) for \jbu is reported with $1\sigma$ errors in parentheses.}
    \label{tab:hst_obs}
    \begin{tabularx}{\columnwidth}{ X X X >{\hsize=2.2cm}X }
        \hline \\
        Date &   Filter & Exposure  & Mag (err)  \\
        \hline \\
        2021-03-02 &  $F275W$  & $4\times450s$ & 25.305 (0.374) \\
        2021-12-06 &  $F350LP$ & $4\times385s$ & 25.685 (0.039) \\
        -          &  $F555W$  & $4\times390s$ & 26.585 (0.112) \\
        -          &  $F814W$  & $4\times390s$ & 25.855 (0.146) \\
        -          &  $F160W$  & $3\times420s$ & 23.744 (0.086) \\
        \hline
    \end{tabularx}
\end{table}

\section{Discussion}

\subsection{Light curve evolution}

\begin{figure}[!ht]
    \centering
    \includegraphics[width= \columnwidth]{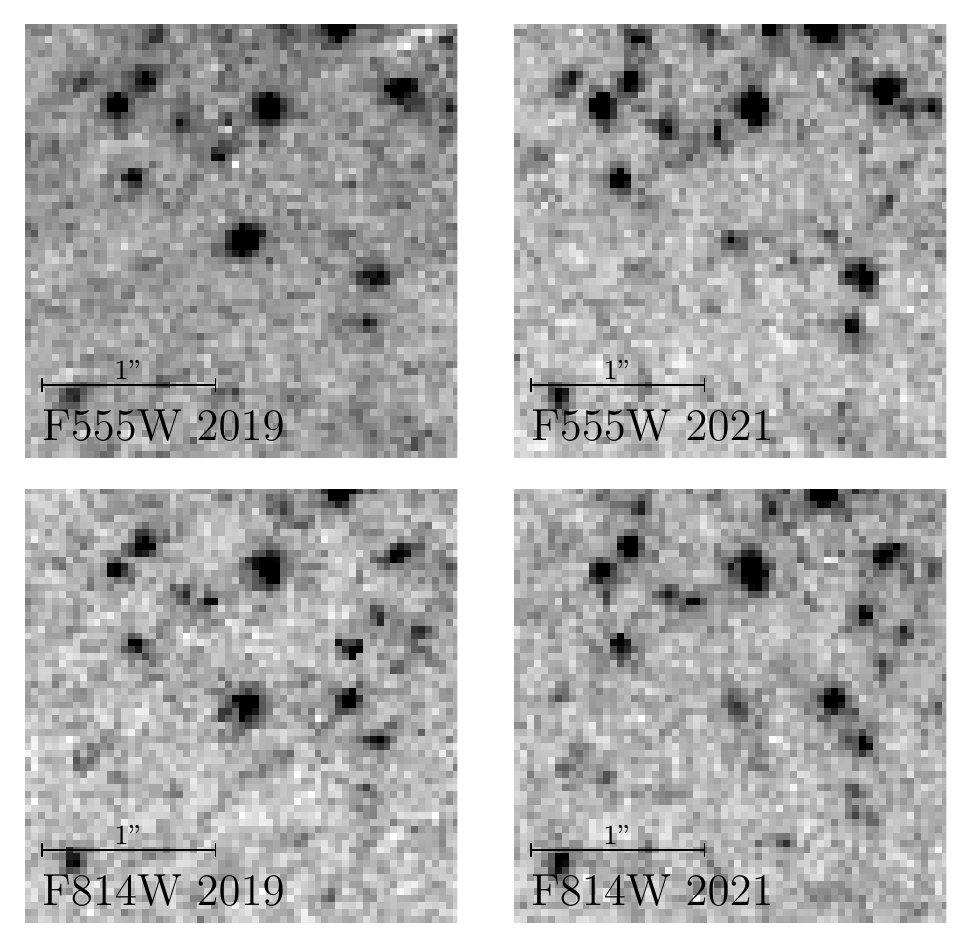}
    \caption{HST cutouts of \jbu in $F555W$ (upper) and $F814W$ (lower) from 2019 (left) and 2021 (right). The position  of \jbu is located centre frame. In each panel, North is pointing upwards and East is pointing left.}
    \label{fig:HST_cutouts}
\end{figure}

As shown in Fig. \ref{fig:HST_cutouts}, \jbu has faded significantly since 2019 and its $F555W$ magnitude is now $\sim2.25$~mag fainter than its minimum value seen in early 2016 \citep{Brennan2021b}. Similar to the recent results from \cite{Smith2022} for \ip, we find a consistent colour (within error) between 2019 and 2021. We measure $(F555W-F814W)_0$ (approximately $V$-$I$) to be $0.47\pm0.04$~mag in 2019; and $0.46\pm0.18$~mag in 2021. Moreover, we find similar decay rates for $F555W$ and $F814W$ between 2019 and 2021 ($\sim991$ days) of  $-0.0027\pm0.0001$ and $-0.0026\pm0.0001$~mag~day$^{-1}$ respectively. These decline rates are almost identical for $F555W$ and $F814W$, and are $~\sim$10 times faster than those observed for \ip \citep{Smith2022} (although still $~\sim$10 times slower than the decline seen in the late time light curve of SN 1987A \citealp[e.g.][]{Woosley1988}). 

\begin{figure*}[!ht]
    \centering
    \includegraphics[width= \textwidth]{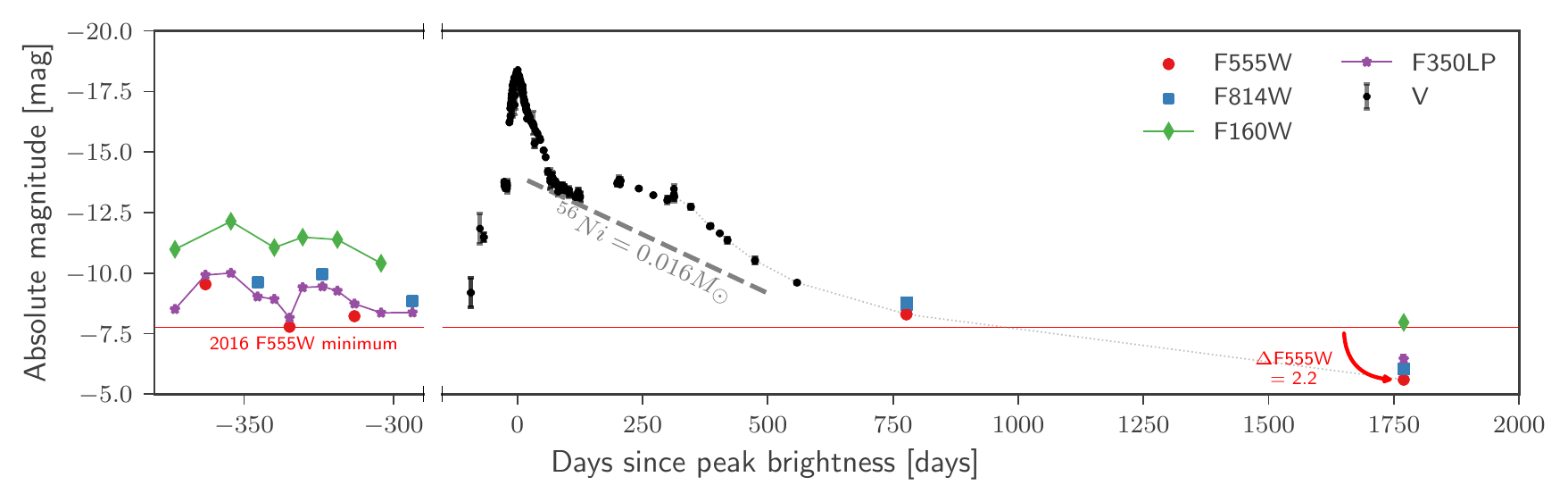}
    \caption{Light curve of \jbu including 
    HST observations from Table. \ref{tab:hst_obs} and \cite{Brennan2021b}, and $V$-band (black markers) observations from \cite{Brennan2021a}. We include the expected luminosity from 0.016~\msun of $^{56}$Ni in the $V$-band \citep{Hamuy2003,Brennan2021b} as the sloped dashed line. We connect points in the late time light curve to visualize the non-linear decline. }
    \label{fig:lightcurve}
\end{figure*}

In Fig. \ref{fig:lightcurve} we plot the $V$-band light curve, as well as the HST observations of \jbu from 2016, 2019, and 2021. From the photometric and spectral features of \jbu, it is obvious that CSM interaction plays a dominant role in the late time evolution \citep{Kilpatrick2018,Brennan2021b}. Amongst other members of the \ip-like transients, unique to \jbu is the re-brightening in the light curve seen after $\sim$+130~d. Notwithstanding this bump, we also see that the decline rate of \jbu from  peak is not linear, with a slower decline between 780 and 1770 days compared to between 250 and 500 days (Fig. \ref{fig:lightcurve}).

The progenitor candidate identified by \cite{Brennan2021b} was a $\sim$22~\msun\ star which exploded as a YHG. One would expect a star with this ZAMS mass to explode as a Type IIP SN, and experience a mass loss rate of $\sim10^{-6}$\msun~yr$^{-1}$ before exploding. Binary evolution can strongly affect the location in the HRD where a star will explode, and we hence investigate whether the light curve of \jbu can be used to confirm or rule out the presence of a binary companion. 

\begin{figure}[!ht]
    \centering
    \includegraphics[width= \columnwidth]{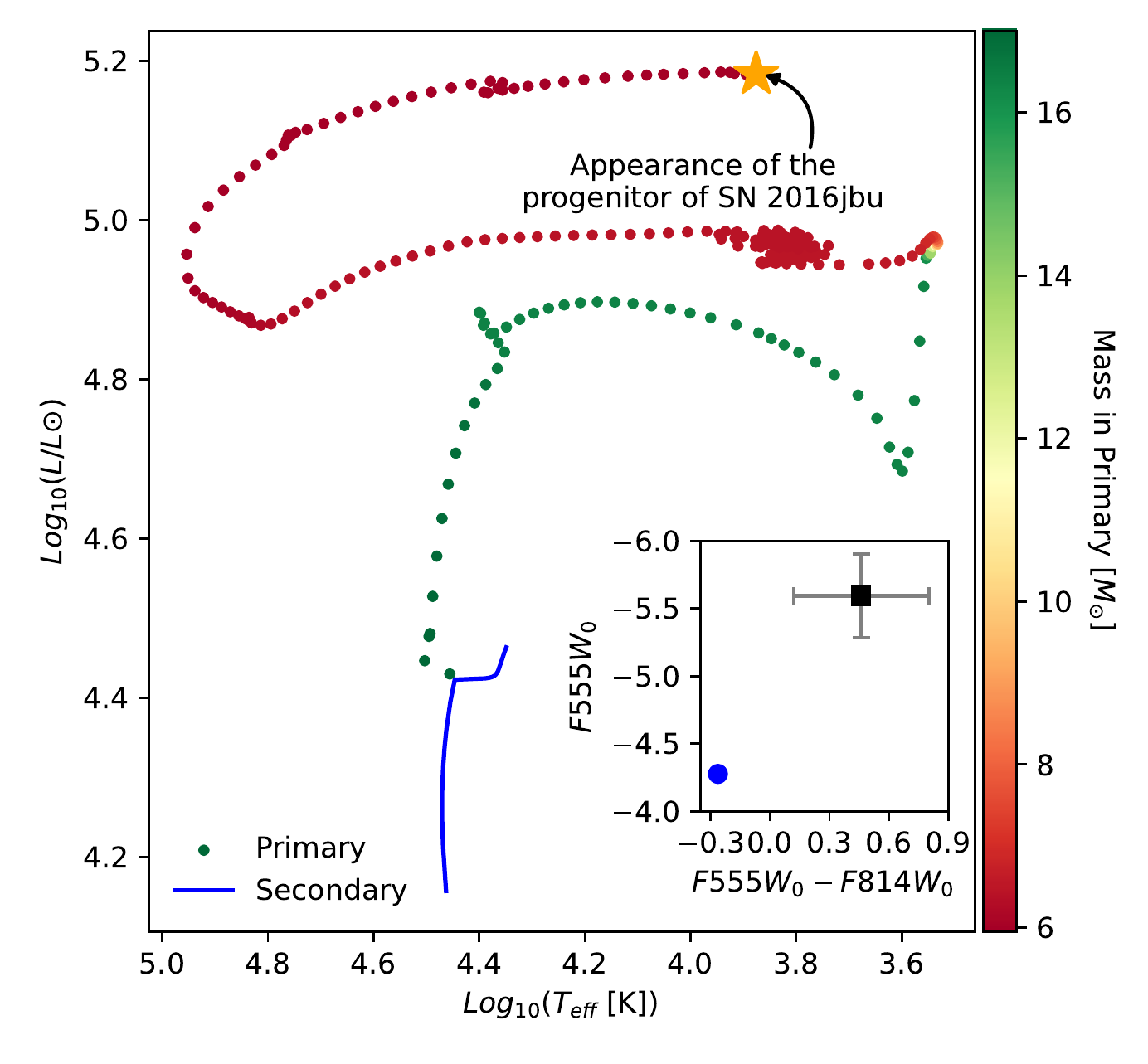}
    \caption{Terminal model matching the progenitor of \jbu from \bpass \citep{Eldridge2017,Stanway2018}. Coloured markers are the primary star ($M_{TAMS}=17$~\msun), assumed to be the progenitor of \jbu. The Blue line is its binary companion ($M_{TAMS}=12$~\msun). We note the evolutionary track for the companion has not reached core-collapse, as the \bpass models terminate when the primary reaches the end of its life. Colour of primary denotes the primaries' mass at each evolutionary time-step, given in the colour bar. Inset shows colour magnitude diagram (CMD) including the 2021 observations of \jbu (black marker) and the colour of the binary companion from the \bpass model during the last evolutionary time-step (blue marker). }
    \label{fig:cmd}
\end{figure}

We compare our 2021 observations to the sample of terminal \bpass \citep{Eldridge2017,Stanway2018} models used in \cite{Brennan2021b} and plot these matching models in Fig.\ref{fig:cmd}. No single star models match the progenitor of \jbu. We find one binary model that matches the 2016 progenitor temperature and luminosity. This model comprises a primary (secondary) with a terminal-age main sequence (TAMS) mass of 17 (12)~\msun. Subsequently, the primary becomes a red supergiant, before loosing most of its mass to its companion and evolving across the HRD to become a hot stripped star. As it crosses the HRD to the blue it begins Helium burning, but then moves back across to the yellow when Carbon burning begins (around $10^4$ years before core-collapse) where it explodes as a yellow hypergiant. When the primary explodes, it still has a significant hydrogen surface fraction ($\sim25\%$), consistent with a Type IIn SN. Encouragingly, \cite{Brennan2021b} explode this model using the {\sc SNEC} code \citep{Morozova2015} and find it can broadly reproduce the \textit{Event B} light curve shape for \jbu. 

We find the companion colour, taken as the last evolutionary time step in the \bpass model, is too blue to match our 2021 observations, $(F555W-F814W)_0\approx+0.46$. Additionally, our $F555W_0$ observations are too bright to be associated with the flux from the companion star alone (the companion magnitude in all bands is fainter than our 2021 measurements). This is consistent with CSM interaction continuing to dominate over any binary companion. Once CSM interaction stops (and if a surviving companion exists), we expect it to be detectable by HST at $F555W$ $\sim27.8$~mag.

\subsection{Modelling the SED}

SN impostors, by definition, do not destroy the progenitor star \citep{VanDyk2000}. We investigate the possibility that the progenitor of \jbu still exists, but is now enshrouded by dust. Using the \dusty radiative transfer code \citep{Ivez1997} we calculate the observed spectral energy distributions (SEDs) for a grid of progenitor models allowing for different configurations of circumstellar dust, see \cite{Brennan2021b} for further details. To remain open to any sort of surviving source, we employ a range of synthetic stellar spectra covering a wide temperature regime, which includes models from {\sc MARCS} \citep{Gustafsson2008}, {\sc PHOENIX} \citep{Husser2013}, {\sc PoWR} \citep{Sander2015}, and {\sc CMFGEN} \citep{Hillier2001}.

\begin{figure}[!ht]
    \centering
     \includegraphics[width= \columnwidth]{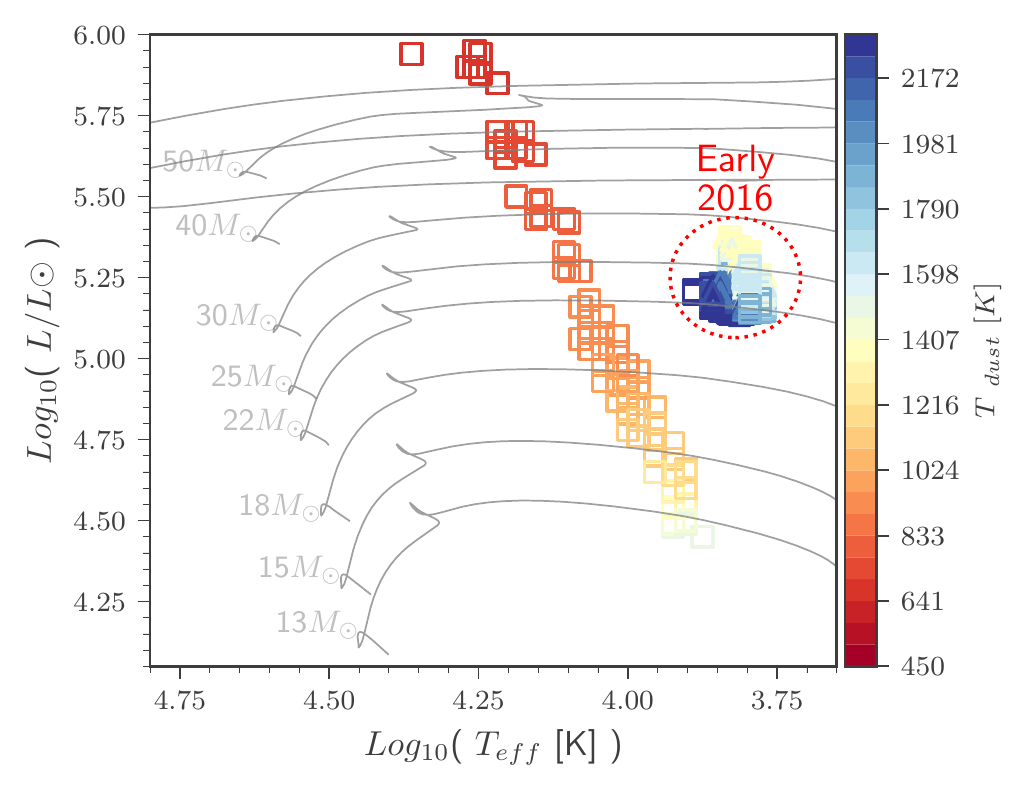}
    \includegraphics[width= \columnwidth]{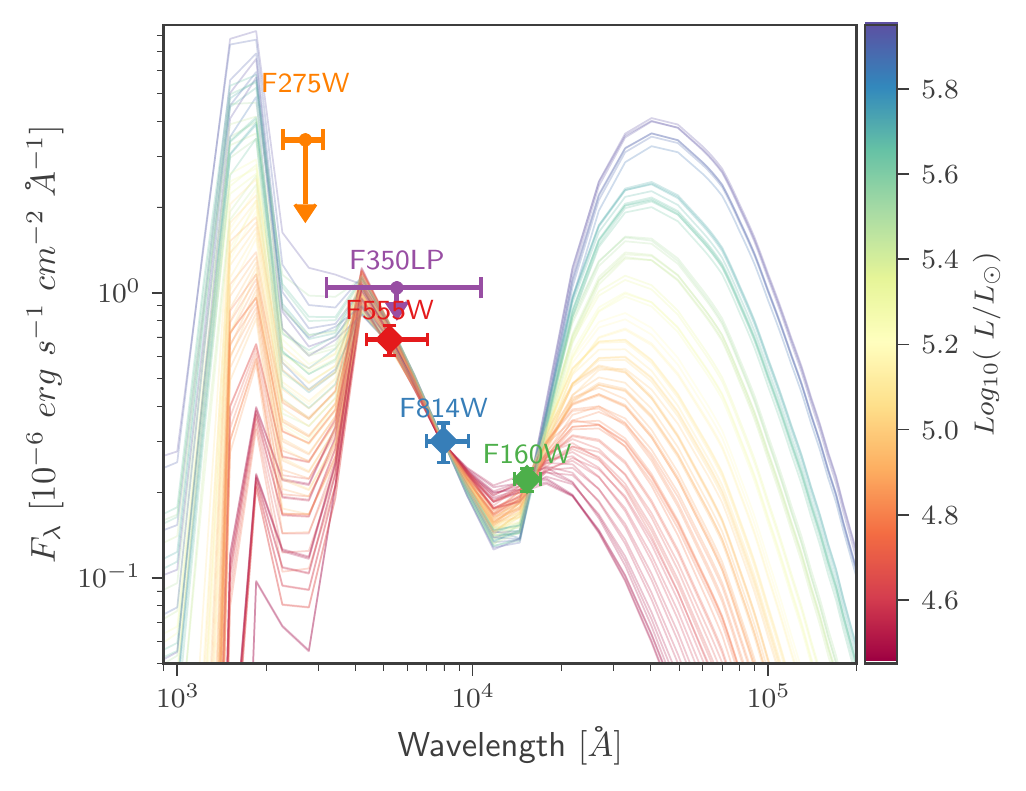}
    \caption{ Upper panel: Hertzsprung-Russell diagram showing model luminosities and effective temperatures. The colour of markers represents the necessary dust temperature required to match the remnant colours. For reference, we overplot single-star evolutionary tracks from the Binary Population and Spectral Synthesis (\bpass) models \citep{Eldridge2017,Stanway2018}.  Lower panel: Matching spectra from \dusty. Markers with error bars represent the extinction corrected flux from the 2021 data given in Table. \ref{tab:hst_obs}. The colour of each spectral energy distribution represents the luminosity of the scaled model. We treat the $F275W$ measurement from March 2021 and the $F350LP$ data from December 2021 as upper limits.}
    \label{fig:matching_spectra}
\end{figure}

For each model, we calculate synthetic $F555W-F814W$ and $F814W-F160W$ colours, and compared them to the foreground extinction corrected colours of the remnant of \jbu, see Table: \ref{tab:hst_obs}\footnote{We exclude $F275W$ and $F350LP$ from our colour matching due to the former being measured in early 2021 and the later likely containing flux from $H\alpha$. We instead use these bands as upper limits.}. Models with colours matching (within measurement errors) \jbu remnant are scaled to match the $F814W_0$ 2021 observation. We then calculate the bolometric luminosity by integrating these scaled SEDs. Matching spectra are plotted in Fig. \ref{fig:matching_spectra}.

We find our 2021 optical measurements can only match a dust-enshrouded source with the luminosity of the progenitor if it is much hotter (upper panel, Fig. \ref{fig:matching_spectra}), but re-radiating much of its flux in the Infra-Red (IR) due to dust (lower panel, Fig. \ref{fig:matching_spectra}). 

The most salient parameter in our modelling is the dust mass. Following \cite{Kochanek2011}, we find these models require $10^{-2} - 10^{-1}$~\msun of graphitic dust to obscure a surviving progenitor\footnote{Assuming an expansion velocity of 5000~\kms, a travel time of 1790 days, and dust opacity of 500~$cm^2~g^{-1}$ \citep{Kochanek2011,Brennan2021b}}. Such dust mass may be expected after $\sim1800$~days post-core-collapse \citep{Wesson2015}, but producing such a large amount of dust at this phase without core-collapse is non-trivial \citep{Kochanek2011}\footnote{ \etacar has $\sim$0.6~\msun of dust, although this is $\sim$150~yrs post eruption \citep{Smith1998} }. Additionally, such a large dust mass would result in a noticeable colour change, which is not seen in \jbu \citep[or \ip, ][]{Smith2022}.

\section{Conclusions}

In this Letter, we have presented and analysed HST/WFC3 images of the source at the position of \jbu taken 5 years after the explosion. 

Our main conclusion is the point source is now significantly fainter ($\sim$2.2mag) than the progenitor \citep{Kilpatrick2018,Brennan2021b}. We find the colour $(F555W-F814W)_0$ has remained roughly constant for $\sim$2 years, suggesting that significant quantities of dust have has not formed. Motivated by the constant colour and non-linear decay seen in \jbu's late-time evolution, we also investigate the possibility that the light curve is fading to reveal a binary companion \citep[e.g.][]{Zhang2004, Kashi2013, Fox2022}. Comparing to models from the \bpass code yields a single matching model. However, the binary companion is too blue and faint to match our 2021 observations, although continued follow-up is needed to confirm this or refute this scenario as the SN fades further.

We attempt to model a surviving progenitor enshrouded by dust using the \dusty code. While our HST photometry can be modelled with a dust-enshrouded star, we cannot tightly constrain the remnant luminosity due to a lack of coverage in the mid-IR (Fig \ref{fig:matching_spectra}). However, any model would require a significant dust mass ($10^{-2} - 10^{-1}$~\msun) to obscure the surviving star and match our optical observations. Such high dust masses are difficult to produce in a non-terminal eruption, and a significant colour change would also be expected.

The above arguments strongly suggest that \jbu was indeed a terminal explosion, and considering the recent observations of \ip \citep{Smith2022} and \bh \citep{Jencson2022}, strongly supports the conclusion that the class of \ip-like transient are indeed genuine - albeit strange - SNe.

However, many questions remain unanswered. What is the underlying mechanism for the Event A/B double peaked light curve seen in the SN2009ip-like transients? How can stellar evolutionary models explain the years-long erratic variability seen prior to explosion in these events? And perhaps most puzzling, why is the class of \ip-like transients relatively homogeneous if they come from such a wider range of progenitors  \citep[e.g. 25 \msun yellow hypergiant for \jbu, and a 60 \msun luminous blue variable for \ip)][]{Smith2010,foley2011,Thone2017, Brennan2021b}?

\begin{acknowledgements}

    S. J. Brennan would like to thank their support from Science Foundation Ireland and the Royal Society (RS-EA/3471).
    
    N.E.R. acknowledges partial support from MIUR, PRIN 2017 (grant 20179ZF5KS), from the Spanish MICINN grant PID2019-108709GB-I00 and FEDER funds, and from the program Unidad de Excelencia María de Maeztu CEX2020-001058-M.

    M. F. is supported by a Royal Society - Science Foundation Ireland University Research Fellowship.
    
    J. D. L. acknowledges support from a UK Research and Innovation Fellowship(MR/T020784/1)
    
    This research is based on observations made with the NASA/ESA Hubble Space Telescope obtained from the Space Telescope Science Institute, which is operated by the Association of Universities for Research in Astronomy, Inc., under NASA contract NAS 5–26555. These observations are associated with program GO-16671.
    
    This research was achieved using the POLLUX database\footnote{\url{http://pollux.oreme.org}} operated at LUPM  (Université Montpellier - CNRS, France with the support of the PNPS and INSU.
    
    This work made use of v2.2.1 of the Binary Population and Spectral Synthesis (\bpass) models as described in \cite{Eldridge2017} and \cite{Stanway2018}.
    
\end{acknowledgements}

\bibliographystyle{aa}
\bibliography{references.bib}

\begin{thebibliography}{45}
\expandafter\ifx\csname natexlab\endcsname\relax\def\natexlab#1{#1}\fi

\bibitem[{{Boian} \& {Groh}(2018)}]{Boian2018}
{Boian}, I. \& {Groh}, J.~H. 2018, \aap, 617, A115

\bibitem[{Brennan {et~al.}(2022{\natexlab{a}})Brennan, Fraser, Johansson,
  Pastorello, Kotak, Stevance, Chen, Eldridge, Bose, Brown, Callis, Cartier,
  Dennefeld, Dong, Duffy, Elias-Rosa, Hosseinzadeh, Hsiao, Kuncarayakti,
  Martin-Carrillo, Monard, Nyholm, Pignata, Sand, Shappee, Smartt, Tucker,
  Wyrzykowski, Abbot, Benetti, Bento, Blondin, Chen, Delgado, Galbany,
  Gromadzki, Gutiérrez, Hanlon, Harrison, Hiramatsu, Hodgkin, Holoien, Howell,
  Inserra, Kankare, Kozłowski, Müller-Bravo, Maguire, McCully, Meintjes,
  Morrell, Nicholl, O’Neill, Pietrukowicz, Poleski, Prieto, Rau, Reichart,
  Schweyer, Shahbandeh, Skowron, Sollerman, Soszyński, Stritzinger,
  Szymański, Tartaglia, Udalski, Ulaczyk, Young, vanLeeuwen, \&
  vanSoelen}]{Brennan2021a}
Brennan, S.~J., Fraser, M., Johansson, J., {et~al.} 2022{\natexlab{a}}, \mnras,
  513, 5642

\bibitem[{Brennan {et~al.}(2022{\natexlab{b}})Brennan, Fraser, Johansson,
  Pastorello, Kotak, Stevance, Chen, Eldridge, Bose, Brown, Callis, Cartier,
  Dennefeld, Dong, Duffy, Elias-Rosa, Hosseinzadeh, Hsiao, Kuncarayakti,
  Martin-Carrillo, Monard, Pignata, Sand, Shappee, Smartt, Tucker, Wyrzykowski,
  Abbot, Benetti, Bento, Blondin, Chen, Delgado, Galbany, Gromadzki,
  Gutiérrez, Hanlon, Harrison, Hiramatsu, Hodgkin, Holoien, Howell, Inserra,
  Kankare, Kozłowski, Müller-Bravo, Maguire, McCully, Meintjes, Morrell,
  Nicholl, O’Neill, Pietrukowicz, Poleski, Prieto, Rau, Reichart, Schweyer,
  Shahbandeh, Skowron, Sollerman, Soszyński, Stritzinger, Szymański,
  Tartaglia, Udalski, Ulaczyk, Young, van Leeuwen, \&
  van Soelen}]{Brennan2021b}
Brennan, S.~J., Fraser, M., Johansson, J., {et~al.} 2022{\natexlab{b}}, \mnras,
  513, 5666

\bibitem[{{Cardelli} {et~al.}(1989){Cardelli}, {Clayton}, \&
  {Mathis}}]{Cardelli1989}
{Cardelli}, J.~A., {Clayton}, G.~C., \& {Mathis}, J.~S. 1989, \apj, 345, 245

\bibitem[{{Dolphin}(2016)}]{2016ascl.soft08013D}
{Dolphin}, A. 2016, {DOLPHOT: Stellar photometry}, Astrophysics Source Code
  Library, record ascl:1608.013

\bibitem[{Eldridge {et~al.}(2017)Eldridge, Stanway, Xiao, McClelland, Taylor,
  Ng, Greis, \& Bray}]{Eldridge2017}
Eldridge, J.~J., Stanway, E.~R., Xiao, L., {et~al.} 2017, Publications of the
  Astronomical Society of Australia, 34

\bibitem[{{Elias-Rosa} {et~al.}(2016){Elias-Rosa}, {Pastorello}, {Benetti},
  {Cappellaro}, {Taubenberger}, {Terreran}, {Fraser}, {Brown}, {Tartaglia},
  {Morales-Garoffolo}, {Harmanen}, {Richardson}, {Artigau}, {Tomasella},
  {Margutti}, {Smartt}, {Dennefeld}, {Turatto}, {Anupama}, {Arbour}, {Berton},
  {Bjorkman}, {Boles}, {Briganti}, {Chornock}, {Ciabattari}, {Cortini},
  {Dimai}, {Gerhartz}, {Itagaki}, {Kotak}, {Mancini}, {Martinelli},
  {Milisavljevic}, {Misra}, {Ochner}, {Patnaude}, {Polshaw}, {Sahu}, \&
  {Zaggia}}]{Elias-Rosa2016}
{Elias-Rosa}, N., {Pastorello}, A., {Benetti}, S., {et~al.} 2016, \mnras, 463,
  3894

\bibitem[{{Filippenko}(1997)}]{Filippenko1997}
{Filippenko}, A.~V. 1997, \araa, 35, 309

\bibitem[{{Foley} {et~al.}(2011){Foley}, {Berger}, {Fox}, {Levesque},
  {Challis}, {Ivans}, {Rhoads}, \& {Soderberg}}]{foley2011}
{Foley}, R.~J., {Berger}, E., {Fox}, O., {et~al.} 2011, \apj, 732, 32

\bibitem[{{Fox} {et~al.}(2015){Fox}, {Smith}, {Ammons}, {Andrews}, {Bostroem},
  {Cenko}, {Clayton}, {Dwek}, {Filippenko}, {Gallagher}, {Kelly}, {Mauerhan},
  {Miller}, \& {Van Dyk}}]{Fox2015}
{Fox}, O.~D., {Smith}, N., {Ammons}, S.~M., {et~al.} 2015, \mnras, 454, 4366

\bibitem[{{Fox} {et~al.}(2022){Fox}, {Van Dyk}, {Williams}, {Drout},
  {Zapartas}, {Smith}, {Milisavljevic}, {Andrews}, {Bostroem}, {Filippenko},
  {Gomez}, {Kelly}, {de Mink}, {Pierel}, {Rest}, {Ryder}, {Sravan}, {Strolger},
  {Wang}, \& {Weil}}]{Fox2022}
{Fox}, O.~D., {Van Dyk}, S.~D., {Williams}, B.~F., {et~al.} 2022, \apjl, 929,
  L15

\bibitem[{{Fraser} {et~al.}(2013){Fraser}, {Inserra}, {Jerkstrand}, {Kotak},
  {Pignata}, {Benetti}, {Botticella}, {Bufano}, {Childress}, {Mattila},
  {Pastorello}, {Smartt}, {Turatto}, {Yuan}, {Anderson}, {Bayliss}, {Bauer},
  {Chen}, {F{\"o}rster Bur{\'o}n}, {Gal-Yam}, {Haislip}, {Knapic}, {Le
  Guillou}, {Marchi}, {Mazzali}, {Molinaro}, {Moore}, {Reichart}, {Smareglia},
  {Smith}, {Sternberg}, {Sullivan}, {Tak{\'a}ts}, {Tucker}, {Valenti}, {Yaron},
  {Young}, \& {Zhou}}]{Fraser2013}
{Fraser}, M., {Inserra}, C., {Jerkstrand}, A., {et~al.} 2013, \mnras, 433, 1312

\bibitem[{{Gustafsson} {et~al.}(2008){Gustafsson}, {Edvardsson}, {Eriksson},
  {J{\o}rgensen}, {Nordlund}, \& {Plez}}]{Gustafsson2008}
{Gustafsson}, B., {Edvardsson}, B., {Eriksson}, K., {et~al.} 2008, \aap, 486,
  951

\bibitem[{{Hamuy}(2003)}]{Hamuy2003}
{Hamuy}, M. 2003, \apj, 582, 905

\bibitem[{{Hillier} \& {Lanz}(2001)}]{Hillier2001}
{Hillier}, D.~J. \& {Lanz}, T. 2001, in Astronomical Society of the Pacific
  Conference Series, Vol. 247, Spectroscopic Challenges of Photoionized
  Plasmas, ed. G.~{Ferland} \& D.~W. {Savin}, 343

\bibitem[{{Hirai} {et~al.}(2021){Hirai}, {Podsiadlowski}, {Owocki},
  {Schneider}, \& {Smith}}]{Hirai2021}
{Hirai}, R., {Podsiadlowski}, P., {Owocki}, S.~P., {Schneider}, F. R.~N., \&
  {Smith}, N. 2021, \mnras, 503, 4276

\bibitem[{{Husser} {et~al.}(2013){Husser}, {Wende-von Berg}, {Dreizler},
  {Homeier}, {Reiners}, {Barman}, \& {Hauschildt}}]{Husser2013}
{Husser}, T.~O., {Wende-von Berg}, S., {Dreizler}, S., {et~al.} 2013, \aap,
  553, A6

\bibitem[{{Ivezic} \& {Elitzur}(1997)}]{Ivez1997}
{Ivezic}, Z. \& {Elitzur}, M. 1997, \mnras, 287, 799

\bibitem[{{Jencson} {et~al.}(2022){Jencson}, {Sand}, {Andrews}, {Smith},
  {Strader}, {Aghakhanloo}, {Pearson}, \& {Valenti}}]{Jencson2022}
{Jencson}, J.~E., {Sand}, D.~J., {Andrews}, J.~E., {et~al.} 2022, arXiv
  e-prints, arXiv:2206.02816

\bibitem[{{Kashi} {et~al.}(2013){Kashi}, {Soker}, \& {Moskovitz}}]{Kashi2013}
{Kashi}, A., {Soker}, N., \& {Moskovitz}, N. 2013, \mnras, 436, 2484

\bibitem[{{Kilpatrick} {et~al.}(2018){Kilpatrick}, {Foley}, {Drout}, {Pan},
  {Panther}, {Coulter}, {Filippenko}, {Marion}, {Piro}, {Rest}, {Seitenzahl},
  {Strampelli}, \& {Wang}}]{Kilpatrick2018}
{Kilpatrick}, C.~D., {Foley}, R.~J., {Drout}, M.~R., {et~al.} 2018, \mnras,
  473, 4805

\bibitem[{{Kochanek}(2011)}]{Kochanek2011}
{Kochanek}, C.~S. 2011, \apj, 743, 73

\bibitem[{{Margutti} {et~al.}(2014){Margutti}, {Milisavljevic}, {Soderberg},
  {Chornock}, {Zauderer}, {Murase}, {Guidorzi}, {Sanders}, {Kuin}, {Fransson},
  {Levesque}, {Chandra}, {Berger}, {Bianco}, {Brown}, {Challis},
  {Chatzopoulos}, {Cheung}, {Choi}, {Chomiuk}, {Chugai}, {Contreras}, {Drout},
  {Fesen}, {Foley}, {Fong}, {Friedman}, {Gall}, {Gehrels}, {Hjorth}, {Hsiao},
  {Kirshner}, {Im}, {Leloudas}, {Lunnan}, {Marion}, {Martin}, {Morrell},
  {Neugent}, {Omodei}, {Phillips}, {Rest}, {Silverman}, {Strader},
  {Stritzinger}, {Szalai}, {Utterback}, {Vinko}, {Wheeler}, {Arnett},
  {Campana}, {Chevalier}, {Ginsburg}, {Kamble}, {Roming}, {Pritchard}, \&
  {Stringfellow}}]{Margutti2014}
{Margutti}, R., {Milisavljevic}, D., {Soderberg}, A.~M., {et~al.} 2014, \apj,
  780, 21

\bibitem[{{Mauerhan} {et~al.}(2013){Mauerhan}, {Smith}, {Filippenko},
  {Blanchard}, {Blanchard}, {Casper}, {Cenko}, {Clubb}, {Cohen}, {Fuller},
  {Li}, \& {Silverman}}]{Mauerhan2013}
{Mauerhan}, J.~C., {Smith}, N., {Filippenko}, A.~V., {et~al.} 2013, \mnras,
  430, 1801

\bibitem[{{Morozova} {et~al.}(2015){Morozova}, {Piro}, {Renzo}, {Ott},
  {Clausen}, {Couch}, {Ellis}, \& {Roberts}}]{Morozova2015}
{Morozova}, V., {Piro}, A.~L., {Renzo}, M., {et~al.} 2015, \apj, 814, 63

\bibitem[{{Nyholm} {et~al.}(2020){Nyholm}, {Sollerman}, {Tartaglia}, {Taddia},
  {Fremling}, {Blagorodnova}, {Filippenko}, {Gal-Yam}, {Howell},
  {Karamehmetoglu}, {Kulkarni}, {Laher}, {Leloudas}, {Masci}, {Kasliwal},
  {Mor{\r{a}}}, {Moriya}, {Ofek}, {Papadogiannakis}, {Quimby}, {Rebbapragada},
  \& {Schulze}}]{Nyholm2020}
{Nyholm}, A., {Sollerman}, J., {Tartaglia}, L., {et~al.} 2020, \aap, 637, A73

\bibitem[{{Pastorello} {et~al.}(2013){Pastorello}, {Cappellaro}, {Inserra},
  {Smartt}, {Pignata}, {Benetti}, {Valenti}, {Fraser}, {Tak{\'a}ts}, {Benitez},
  {Botticella}, {Brimacombe}, {Bufano}, {Cellier-Holzem}, {Costado}, {Cupani},
  {Curtis}, {Elias-Rosa}, {Ergon}, {Fynbo}, {Hambsch}, {Hamuy}, {Harutyunyan},
  {Ivarson}, {Kankare}, {Martin}, {Kotak}, {LaCluyze}, {Maguire}, {Mattila},
  {Maza}, {McCrum}, {Miluzio}, {Norgaard-Nielsen}, {Nysewander}, {Ochner},
  {Pan}, {Pumo}, {Reichart}, {Tan}, {Taubenberger}, {Tomasella}, {Turatto}, \&
  {Wright}}]{Pastorello2013}
{Pastorello}, A., {Cappellaro}, E., {Inserra}, C., {et~al.} 2013, \apj, 767, 1

\bibitem[{{Pastorello} {et~al.}(2018){Pastorello}, {Kochanek}, {Fraser},
  {Dong}, {Elias-Rosa}, {Filippenko}, {Benetti}, {Cappellaro}, {Tomasella},
  {Drake}, {Harmanen}, {Reynolds}, {Shappee}, {Smartt}, {Chambers}, {Huber},
  {Smith}, {Stanek}, {Christensen}, {Denneau}, {Djorgovski}, {Flewelling},
  {Gall}, {Gal-Yam}, {Geier}, {Heinze}, {Holoien}, {Isern}, {Kangas},
  {Kankare}, {Koff}, {Llapasset}, {Lowe}, {Lundqvist}, {Magnier}, {Mattila},
  {Morales-Garoffolo}, {Mutel}, {Nicolas}, {Ochner}, {Ofek}, {Prosperi},
  {Rest}, {Sano}, {Stalder}, {Stritzinger}, {Taddia}, {Terreran}, {Tonry},
  {Wainscoat}, {Waters}, {Weiland}, {Willman}, {Young}, \&
  {Zheng}}]{Pastorello2018}
{Pastorello}, A., {Kochanek}, C.~S., {Fraser}, M., {et~al.} 2018, \mnras, 474,
  197

\bibitem[{{Reilly} {et~al.}(2017){Reilly}, {Maund}, {Baade}, {Wheeler},
  {H{\"o}flich}, {Spyromilio}, {Patat}, \& {Wang}}]{Reilly2017}
{Reilly}, E., {Maund}, J.~R., {Baade}, D., {et~al.} 2017, \mnras, 470, 1491

\bibitem[{{Sander} {et~al.}(2015){Sander}, {Shenar}, {Hainich},
  {G{\'\i}menez-Garc{\'\i}a}, {Todt}, \& {Hamann}}]{Sander2015}
{Sander}, A., {Shenar}, T., {Hainich}, R., {et~al.} 2015, \aap, 577, A13

\bibitem[{{Schlegel}(1990)}]{Schlegel1990}
{Schlegel}, E.~M. 1990, \mnras, 244, 269

\bibitem[{{Smith} {et~al.}(2022){Smith}, {Andrews}, {Filippenko}, {Fox},
  {Mauerhan}, \& {Van Dyk}}]{Smith2022}
{Smith}, N., {Andrews}, J.~E., {Filippenko}, A.~V., {et~al.} 2022, arXiv
  e-prints, arXiv:2205.02896

\bibitem[{Smith {et~al.}(1998)Smith, Gehrz, \& Krautter}]{Smith1998}
Smith, N., Gehrz, R.~D., \& Krautter, J. 1998, The Astronomical Journal, 116,
  1332

\bibitem[{{Smith} {et~al.}(2014){Smith}, {Mauerhan}, \& {Prieto}}]{Smith2014}
{Smith}, N., {Mauerhan}, J.~C., \& {Prieto}, J.~L. 2014, \mnras, 438, 1191

\bibitem[{{Smith} {et~al.}(2010){Smith}, {Miller}, {Li}, {Filippenko},
  {Silverman}, {Howard}, {Nugent}, {Marcy}, {Bloom}, {Ghez}, {Lu}, {Yelda},
  {Bernstein}, \& {Colucci}}]{Smith2010}
{Smith}, N., {Miller}, A., {Li}, W., {et~al.} 2010, \aj, 139, 1451

\bibitem[{Stanway \& Eldridge(2018)}]{Stanway2018}
Stanway, E.~R. \& Eldridge, J.~J. 2018, \mnras, 479, 75–93

\bibitem[{{Tartaglia} {et~al.}(2016){Tartaglia}, {Pastorello}, {Sullivan},
  {Baltay}, {Rabinowitz}, {Nugent}, {Drake}, {Djorgovski}, {Gal-Yam},
  {Fabrika}, {Barsukova}, {Goranskij}, {Valeev}, {Fatkhullin}, {Schulze},
  {Mehner}, {Bauer}, {Taubenberger}, {Nordin}, {Valenti}, {Howell}, {Benetti},
  {Cappellaro}, {Fasano}, {Elias-Rosa}, {Barbieri}, {Bettoni}, {Harutyunyan},
  {Kangas}, {Kankare}, {Martin}, {Mattila}, {Morales-Garoffolo}, {Ochner},
  {Rebbapragada}, {Terreran}, {Tomasella}, {Turatto}, {Verroi}, \&
  {Wo{\'z}niak}}]{Tartaglia2016}
{Tartaglia}, L., {Pastorello}, A., {Sullivan}, M., {et~al.} 2016, \mnras, 459,
  1039

\bibitem[{{Th{\"o}ne} {et~al.}(2017){Th{\"o}ne}, {de Ugarte Postigo}, \&
  {Leloudas}}]{Thone2017}
{Th{\"o}ne}, C.~C., {de Ugarte Postigo}, A., \& {Leloudas}, G. 2017, in The
  Lives and Death-Throes of Massive Stars, ed. J.~J. {Eldridge}, J.~C. {Bray},
  L.~A.~S. {McClelland}, \& L.~{Xiao}, Vol. 329, 44--48

\bibitem[{{Van Dyk} \& {Matheson}(2012)}]{VanDyk2012}
{Van Dyk}, S.~D. \& {Matheson}, T. 2012, in Astrophysics and Space Science
  Library, Vol. 384, Eta Carinae and the Supernova Impostors, ed. K.~{Davidson}
  \& R.~M. {Humphreys}, 249

\bibitem[{{Van Dyk} {et~al.}(2000){Van Dyk}, {Peng}, {King}, {Filippenko},
  {Treffers}, {Li}, \& {Richmond}}]{VanDyk2000}
{Van Dyk}, S.~D., {Peng}, C.~Y., {King}, J.~Y., {et~al.} 2000, \pasp, 112, 1532

\bibitem[{{Wesson} {et~al.}(2015){Wesson}, {Barlow}, {Matsuura}, \&
  {Ercolano}}]{Wesson2015}
{Wesson}, R., {Barlow}, M.~J., {Matsuura}, M., \& {Ercolano}, B. 2015, \mnras,
  446, 2089

\bibitem[{{Woosley}(1988)}]{Woosley1988}
{Woosley}, S.~E. 1988, \apj, 330, 218

\bibitem[{{Woosley} {et~al.}(2002){Woosley}, {Heger}, \&
  {Weaver}}]{Woosley2002}
{Woosley}, S.~E., {Heger}, A., \& {Weaver}, T.~A. 2002, Reviews of Modern
  Physics, 74, 1015

\bibitem[{{Yaron} {et~al.}(2017){Yaron}, {Perley}, {Gal-Yam}, {Groh}, {Horesh},
  {Ofek}, {Kulkarni}, {Sollerman}, {Fransson}, {Rubin}, {Szabo}, {Sapir},
  {Taddia}, {Cenko}, {Valenti}, {Arcavi}, {Howell}, {Kasliwal}, {Vreeswijk},
  {Khazov}, {Fox}, {Cao}, {Gnat}, {Kelly}, {Nugent}, {Filippenko}, {Laher},
  {Wozniak}, {Lee}, {Rebbapragada}, {Maguire}, {Sullivan}, \&
  {Soumagnac}}]{Yaron2017}
{Yaron}, O., {Perley}, D.~A., {Gal-Yam}, A., {et~al.} 2017, Nature Physics, 13,
  510

\bibitem[{{Zhang} {et~al.}(2004){Zhang}, {Wang}, {Zhou}, {Li}, {Ma}, {Jiang},
  \& {Li}}]{Zhang2004}
{Zhang}, T., {Wang}, X., {Zhou}, X., {et~al.} 2004, \aj, 128, 1857

\end{thebibliography}




\end{document}